\documentclass[prl,aps,amsfonts,amssymb,floats,
twocolumn,showpacs]{revtex4}
\usepackage{epsfig}
\usepackage{graphicx}

\begin{document}

\title{Critical Level-Spacing Distribution for General Boundary Conditions}

\author{S.N. Evangelou\footnote{e-mail: sevagel@cc.uoi.gr,
permanent address: Physics Department, University of Ioannina,
Ioannina 45110, Greece}}

\affiliation{Max-Planck-Institut for the Physics of Complex Systems,
Noethnitzer Strasse 38, D-01187 Dresden, Germany}
 
\begin{abstract}
It is believed that the semi-Poisson function
$P(S)=4S\exp(-2S)$ describes the normalized distribution of the
nearest level-spacings $S$  for critical energy levels at the
Anderson metal-insulator transition from quantum chaos to integrability, 
after an average over four obvious boundary conditions (BC) 
is taken (Braun {\it et} {\it al} \cite{1}). 
In order to check whether the semi-Poisson 
is the correct universal distribution at
criticality we numerically compute it by integrating 
over all possible boundary conditions. We find that
although $P(S)$ describes very well the main part of the obtained
critical distribution small differences exist particularly in the large $S$
tail. The simpler crossover between the integrable ballistic and 
localized limits is shown to be universally characterized 
by a Gaussian-like $P(S)$ distribution instead.
\end{abstract}


\maketitle

\narrowtext
 
 
\par 
\medskip 
Chaos in quantum systems is known to appear in the level 
statistics of the spectral fluctuations for stationary energy 
levels obtained from Hamiltonians defined in appropriate Hilbert 
spaces \cite{2}. In the quantum chaotic regime the loss of most 
symmetries is manifested via the appearance of level-repulsion 
between nearest neighbor energy levels which implies a highly 
correlated spectrum \cite{3}. Integrable systems display, instead, 
uncorrelated spectra with random levels which allow spectral 
degeneracies as unambiguous signatures of high symmetry. The 
solution of Schr$\ddot{o}$dinger equation for billiards having 
zero potential inside and infinite outside a bounded domain
manifests integrable or chaotic  quantum behavior depending on
the shape of the domain boundary.  This quantum crossover 
corresponds to the classical transition from integrability
to chaos  in billiards \cite{4}.  
Wave chaotic behavior has recently been 
observed experimentally for optical billiards \cite{5}. 

\par
\medskip 
Disorder in quantum systems, e.g. a random potential
for electrons in a lattice, is known to cause an Anderson 
metal-insulator transition from extended to localized wave 
functions \cite{6}.  
This is a true quantum phase transition from
chaos to integrability with the extended wave functions 
of weakly disordered metals corresponding to 
quantum chaos and the localized wave functions  
of strongly disordered insulators to integrability. 
By adding weak disorder to a perfect system a  
ballistic to chaotic quantum crossover also 
occurs, similarly to quantum billiards.  Only
for stronger disorder one can obtain the metal-insulator 
transition from diffusive extended to integrable 
localized states.  
One dimension is exceptional in this respect
since no intermediate quantum chaotic behavior 
can be seen with a crosover straight from ballistic 
to localized behavior.
 
\par
\medskip
The critical region at the metal-insulator transition
between diffusion and localization  has also been explored 
for level statistics in terms of the nearest neighbor level-spacing
distribution $P(S)$ \cite{7,8}. 
The disordered problem has certain advantages, such as  the existence
of many energy levels at the transition and a statistical ensemble 
over disorder which complements the usual energy ensemble.
The corresponding critical $P(S)$ 
turns out to be a scale-invariant hybrid of chaotic Wigner-like linear
small $S$ and an integrable Poisson-like exponential for 
large $S$. The main surprise, however, in
these studies was the generic sensitivity of the obtained critical
level statistics to boundary conditions (BC) \cite{1}. This is now
well understood since it simply reflects the nature of critical wave 
functions, which are in between extended and localized, described 
by scale-invariant fractal distributions. Their filamented structure,
closer to one-dimensional, brings little overlap at the boundaries 
so it is accompanied by extreme sensitivity in boundary conditions. 
This is the reason for the corresponding generic dependence 
of critical level statistics to BC. In this Letter we generalize 
the argument of Ref. \cite{1} by including many possible BC 
via the introduction of all Aharonov-Bohm fluxes 
and perform an integration over them.
 
\par
\medskip
Our aim is to compare the obtained
level-spacing distribution at the metal-insulator transition
with the simple semi-Poisson form
$P(S)=4 S\exp(-2S)$ \cite{9}. Although a similar hybrid
distribution describes, at least partially, the level-spacings in
various metal-insulator transitions in disordered systems
including a system with spin-orbit coupling \cite{10}, 
the precise semi-Poisson form often
fails to describe criticality and is not widely accepted 
as the universal critical distribution. 
We ask the question whether a semi-Poisson might be
the appropriate critical distribution to describe averages over
all possible BC. For this purpose we set up and diagonalize
Hamiltonians with disorder defined for cubic lattice clusters 
in three dimensions (3D) with many possible
BC and integrate over all of them. We shall also consider 
the simpler quantum crossover from the ballistic to localized phase 
in 1D disordered systems. In this case we obtain a universal 
Gaussian-like distribution at the crossover where the scaling
parameter of the localization length $\xi$ over size $L$
is fixed to unity.

\par
\medskip
Quantum chaotic systems
are well described by random matrix ensembles via Random
Matrix Theory (RMT) \cite {2,3,4}. 
The level-spacing distribution for small
$S$ involves the level-repulsion condition $P(S)\propto
S^{\beta}$, where the so-called universality class index
$\beta=1,2,4$ is determined from the remaining fundamental symmetries
of quantum chaotic systems. In the presence of time reversal
invariance one obtains the orthogonal universality class with
$\beta=1$.   An applied magnetic field which breaks
the time-reversal invariance leads to
unitary universality class with $\beta=2$ while
spin-orbit coupling with time-reversal to
symplectic universality class with $\beta=4$. The critical $P(S)$
is also not independent of the remaining symmetries of a chaotic system
either so that appropriate expressions for the corresponding critical
$P(S)$ in different universality classes can be found \cite{10}. 
In the rest of the paper we shall rely on
numerical tools in order to explore level statistics at
criticality. Our aim is to answer the previously posed question
whether the semi-Poisson distribution appears 
after averaging over all possible BC.
 
\par 
\medskip 
We obtain spectra and their fluctuations in cubic 3D lattices
with disorder.  We have used standard diagonalization algorithms 
to compute eigenvalues for complex Hermitian sparse random matrices 
of size $L^{3}$ which include flux.  Many
possible BC are considered by adding fluxes
in all directions. More details of the numerical BC
integration method can be found in \cite{11} where it was used for
Hubbard models to describe interacting electrons.
In the quantum coherent chaotic metallic regime 
the results for the eigenvalue
level-spacing distribution $P(S)$ are described by the Wigner
surmise \cite{3}. 
In the insulating integrable regime quantum coherence is
lost since the localized states do not communicate with each other 
located in random positions, so that one obtains the
random Poisson distribution. At criticality after integration
over all BC we obtain a critical $P(S)$ which is closely described by 
semi-Poisson. However, we also find rather small deviations, 
particularly in the large $S$ tail of the distribution.


\par
\medskip
We consider a tight-binding model with site randomness defined on
a $L\times L\times L$ finite lattice described by the Hamiltonian
with potential energy and kinetic energy terms
\begin{eqnarray}
H=\sum_{n}V_{n}c^{\dag }_nc_m- \sum_{\langle nm \rangle }(c^{\dag
}_nc_m+c^{\dag }_mc_n).
\end{eqnarray}
The sum is taken over all lattice sites $n$ and all bonds ${\langle nm
\rangle }$ where $n,m$ denote nearest neighbor lattice sites,
$c_n$ is an annihilation operator of an electron
on site $n$,  $c^{\dag}_n$ is a corresponding 
creation operator of an electron and $V_{n}$ is the random
site potential, a real independent random variable 
satisfying a box distribution $P(V_{n}) = \frac{1}{W}, 
\text{ for }-\frac{W}{2}
\leq V_{n} \leq \frac{W}{2}$ of  zero mean and width $W$ which
measures the strength of disorder. 
We impose the generalized periodic boundary conditions on the 
wave functions $\psi(x,y,z)$ in all directions
\begin{eqnarray}
\psi(x+L,y,z)=e^{i\alpha_{x}}\psi(x,y,z) \\ \nonumber
\psi(x,y+L,z)=e^{i\alpha_{y}}\psi(x,y,z) \\
\psi(x,y,z+L)=e^{i\alpha_{z}}\psi(x,y,z)\nonumber,
\end{eqnarray}
which is equivalent to
applying Aharonov-Bohm magnetic flux ${\vec{\alpha}}$ with
components $\left(\alpha_{x},\alpha_{y},\alpha_{z}\right)$
varying in $[0,2\pi)$.

\par
\medskip
For the zero disorder $W=0$ ballistic limit Eq. (1) has eigenvalues
$\epsilon_{j_{x},j_{y},j_{z}}(k_x, k_y,
k_z)=-2(\cos{k_{x}}+\cos{k_{y}}+\cos{k_{z}})$ and plane wave
eigenstates $\psi_{j_{x},j_{y},j_{z}}(x,y,z)=L^{-3/2}
\exp(-i(k_{x}x+k_{y}y+k_{z}z)$, with the positions of the
$\vec{k}$-vectors determined by the BC of Eq. (2) via
\begin{eqnarray}
k_{x}={\frac {\alpha_{x}+2\pi j_{x}} {L}}, ~~j_x=0, ... , L-1\\ \nonumber
k_{y}={\frac {\alpha_{y}+2\pi j_{y}} {L}}, ~~j_y=0, ... , L-1  \\
k_{z}={\frac {\alpha_{z}+2\pi j_{z}} {L}}, ~~j_z=0, ... , L-1  \nonumber.
\end{eqnarray}
For a given set of BC the allowed values of $k_{x}, k_{y}, k_{z}$ 
form a rigid grid of $L\times L\times L$ points in the Brillouin zone. 
By varying the $\alpha_{x,y,z}$'s
within $[0,2\pi)$ each $k$-point in the Brillouin zone
shifts to cover its own box so that the boxes of all points
partition exactly the Brillouin zone \cite{11}.
When $\alpha_{x,y,z}=0$ or $\pi$ the added flux is equivalent to
periodic, antiperiodic BC in the appropriate $x, y, z$ direction,
respectively.
For critical disorder $W_{c}=16.4$ \cite{7,8}  Eq. (1) has
eigenvalues $\epsilon_{j_{x},j_{y},j_{z}}$ and fractal eigenstates
$\psi_{j_{x},j_{y},j_{z}}(x,y,z)$. For any disorder $W$ and
size $L$ the eigenvalues are obtained by numerical diagonalization.



\par
\medskip
To characterize spectral fluctuations we compute the
nearest level-spacing $P(S)$ distribution function. First, we
make the local mean level-spacing $\Delta(E) \propto 1/\rho(E)$
constant equal to one where $\rho$ is the mean density of states,
by ``unfolding" the spectrum via local
rescaling of the energy. For this purpose one needs the
local average level-spacing $\Delta(E)$ which can be obtained over many
levels around $E$. Alternatively, the $i$th ``unfolded" energy level
${\cal {E}}_{i}={\cal {N}}_{av}(E_{i})$
can be obtained from the averaged integrated spectral density ${\cal
{N}}_{av}(E)$,
leading to average nearest level-spacing $\langle S\rangle
=\langle {\cal E}_{i}-{\cal E}_{i-1}\rangle=1$.
In the studied energy regime around the band center
the results did not change much  with or without a proper unfolding 
since the mean density of states is almost constant \cite{7}.

\begin{figure}
\includegraphics[width=8.75cm]{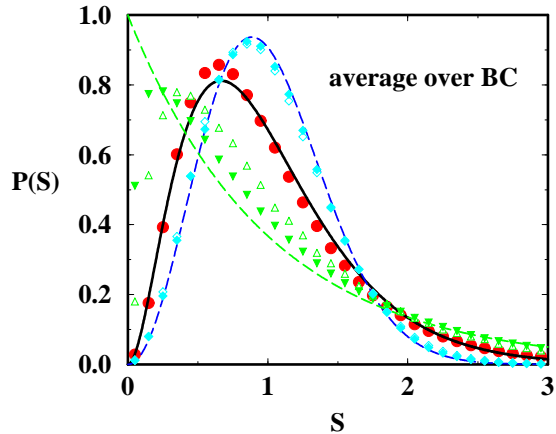}
\caption{ Metal-insulator transition. The level-spacing
distribution function $P(S)$ is shown in 3D disordered
systems for the integrable, critcal and chaotic regimes
for $10$ random configurations of linear size $L=4, 10$ 
by integrating over 1000
possible boundary conditions, including periodic and antiperiodic,
corresponding to 1000 points in $\vec{k}$-space making up a total
number of a few million eigenvalues. As the system size increases
the data for the metallic regime with $W=10$ and the insulating 
regime with $W=30$ approach the dashed lines on the right and
left of the figure, which are the Wigner surmise $P(S)=(\pi/2)S
\exp(-(\pi/4)S^{2})$ and the Poisson law $P(S)=\exp(-S)$,
respectively. The filled dots of the computed scale-invariant
distribution at critical disorder strength $W_{c}=16.4$ are
compared with the semi-Poisson distribution $P(S)=4S \exp(-2S)$
denoted by the solid line.}
\end{figure}

\par
\medskip
In Fig. 1 we present the obtained level-spacing distribution
$P(S)$ in 3D for various sizes $L$ with disorder $W$
by integrating over 1000 boundary conditions (BC). 
We have used the majority (about a third of the total) 
of energy levels centered around the band center 
ignoring only those at band tails.
For a given finite size $L$ the results of disorder
strength $W$ corresponding to localization length $\xi$
lie in-between the Wigner surmise ($L\to \infty$ asymptotic 
for extended states) and the Poisson distribution ($L\to \infty$ 
asymptotic for localized states) \cite{7}.  Only for 
large enough $L \gg \xi$ the data eventually approach the
Poisson curve which for $W> W_{c}$ indicates only localized states
in 3D. For  $W< W_{c}$  the data lie 
close to Wigner surmise which is also quickly
approached for large size. 
Our results of  Fig. 1 below the critical point 
($W=10< W_{c}$) demonstrate a $P(S)$ which approaches  Wigner 
and above the critical point ($W=30> W_{c}$)
a $P(S)$ which approaches Poisson, respectively. 
At $W=W_{c}$ the obtained scale-invariant 
critical $P(S)$  averaged over BC
has Wigner-like small spacing behavior 
and Poisson-like large spacing behavior. 
It is roughly described by the semi-Poisson curve \cite{1} 
although some differences become obvious, particularly
in the tail of the distribution 
as seen in the semi-log plot of
Fig. 2. These deviations also seem to violate 
the scale-invariance of the critical 
curve, since they slightly increase 
by increasing $L$.
However, in this part of the distribution $P(S)$ is very low
and it is much more likely that numerical accuracy of the data 
is lost.
 
\begin{figure}
\includegraphics[width=6.00cm]{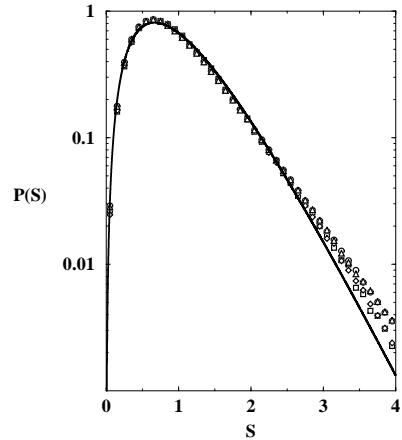}
\caption{A linear-log plot of the critical $P(S)$ shown in Fig. 1
where prominent deviations from the semi-Poisson 
can be seen in the large-$E$ tail.}
\end{figure}

\medskip
\par
We have also examined the simpler ballistic to localization 
crossover.  It occurs in $1D$ disordered systems
between two integrable limits, one purely quantum where
the kinetic energy term of $H$ in Eq. (1) dominates and the other
classical characterized the fluctuating random potential
energy term of Eq. (1). 
In the ballistic regime one asymptotically reaches 
a simple $\delta$-function $P(S)$ while in the localized regime 
the Poisson distribution. Our results taken at the band center
$E\approx 0$ are displayed in Fig. 3. Since the
localization length is easily computed 
with the anomalous perturbational result
$\xi \approx 105.24W^{-2}$ at $E=0$ \cite{12,13}, 
different from usual perturbation $\xi \approx 96W^{-2}
\sqrt{1-(E/2)^{2}}$ valid at most energies $E$.
We vary  $W$  for many random configurations
which changes $\xi$ by considering appropriate sizes 
$L$ to keep the scaling ratio $\xi/L$ fixed. 
In the ballistic  limit  we find that 
a delta-function  $P(S)$ is approached
for $\xi/L\gg1$ while in
the localized limit Poisson is reached for
$\xi/L \ll1$. For $\xi/L=1/5$ we obtain a $P(S)$ 
closer to Poisson shown in Fig. 3(b).
In Fig. 3(a) we display the obtained 
crossover distribution 
at the fixed ratio $\xi/L=1$ for various 
$\xi$'s (varying $W$) and $L$'s.
We observe that the scale-invariant
crossover distribution is similar
to a simple Gaussian.  Of course differences exist, such as 
the requirement of positive $S$ which does not 
allow the existence of Gaussian tails moving in the left-hand
side. We emphasise that the computed $P(S)$
measures fluctuations over the disorder ensemble
precisely at the band center, rather than fluctuations 
in the energy domain studied in billiards.

\begin{figure}
\includegraphics[width=8cm]{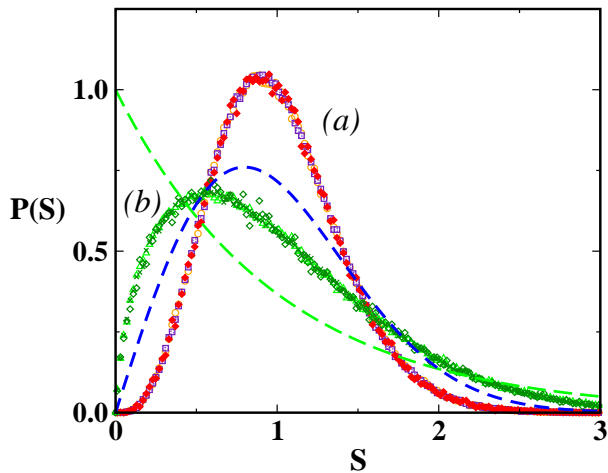}
\caption{ Ballistic to localized crossover. 
The level-spacing distribution function $P(S)$ 
in 1D disordered systems is shown to
depend on the scaling parameter $\xi/L$, where $\xi$ 
is the localization length and $L$ the size. {\bf (a)}
In the crossover regime between 
ballistic and localized states having fixed $\xi/L=1$
from diagonalization of $100000$,
$500000$ and $1000000$ random matrixes of size $L=500, 1000, 5000$
for energies close to $E=0$ with appropriate disorder values $W$.
{\bf (b)} In the localized regime having $\xi/L=1/5$
from diagonalization of 500,000 random matrices of $L=500,1000, 5000$.
In the ballistic regime $\xi/L>1$ the $P(S)$ (not shown)
rapidly approaches a delta function centered around the mean $S=1$.}
\end{figure}

 
\par
\medskip
We have presented numerical results for 
the level-spacing distribution $P(S)$
of non-interacting electrons in the presence
of potential energy fluctuations
at the Anderson metal-insulator transition.
The critical behavior characterized by multifractal 
eigenstates leads to a generic sensitivity to BC \cite{1}. 
In the present work imposed Aharonov-Bohm fluxes 
are equivalent to considering many BC.
For example, a continuous set of BC in 1D  
leads to the equivalent problem of studying
the non-overlapping mini-band widths \cite{14}. 
The AB-flux controlled crossover from orthogonal to unitary studied
in \cite{15} has also revealed sensitivity
to the imposed AB fluxes.
Although such sampling over the Brillouin zone
via the Aharonov-Bohm fluxes breaks time-reversal
invariance, by changing the universality class from orthogonal to
unitary, the obtained results remain valid for the
orthogonal case.

\par
\medskip
In summary,  the $P(S)$ distribution
for disordered systems  is computed
to test the semi-Poisson form at criticality.  
The obtained results complement previous
studies for understanding the metal-insulator
transition. Our main finding is that the corresponding
critical statistics is reasonably well described by
the semi-Poisson function after an average over many BC
is taken.  Small but rather persistent deviations occur 
mostly in the difficult to reach large $S$ tail,
possibly due to numerical artifacts arising from limited accuracy.
The spectral fluctuations studied here could help to understand 
various properties of current-currying states in mesoscopic systems
and might also suggest a critical RMT.
A disorder system is seen to progress from ballistic 
extended (periodic integrable) to diffusive (chaotic) and 
finally to localized (random integrable) regimes.
The ballistic regime is dominated by quantum correlations
and the localized regime by random potential energy fluctuations,
respectively.  The intermediate quantum chaotic
regime is reached either by shape peculiarities
in billiards or for moderate disordered potential 
in solids. 
We add the remark that the Gaussian-like $P(S)$ observed 
for the addition spectra of the conductance peaks in Coulomb blockade
\cite{16}, if it could be interpreted within a non-interacting
electron framework, would suggest an analogy to the obtained
crossover between two integrable regimes with and without disorder.

I like to thank H. Schomerus for a critical reading of the manuscript.


\begin{thebibliography}{99}
 
\bibitem{1}  D. Braun, G. Montambaux, and M. Pascaud,
Phys. Rev. Lett. {\bf 81}, 1062 (1998).

\bibitem{2} T. Guhr, A. M$\ddot{u}$ller-Groeling, and H.
Weidenm$\ddot{u}$ller, Phys. Rep. {\bf 229}, 189-425 (1998).

\bibitem{3} O. Bohigas, M.J. Giannoni and C. Schmit,
Phys. Rev. Lett. {\bf 52}, 1 (1984), see also O. Bohigas, in {\it
Chaos and Quantum Physics}, Proceedings of Les Houches Summmer
School, edited by M.J. Giannoni, A. Voros, and J. Zinn-Justin,
Elsevier, New York (1991).

\bibitem{4} H.-J. St$\ddot{o}$ckmann, {\it ``Quantum Chaos:
an Intorduction"}, Cambridge University Press, New York, 1999.

\bibitem{5}V. Milner, J.L. Hansen, W.C. Campbell, M.G. Raizen,
Phys. Rev. Lett. {\bf 86}, 1514 (2001); N. Friedman, A. Kaplan, D.
Carasso and N. Davidson, Phys. Rev. Lett. {\bf 86}, 1518 (2001).

\bibitem{6} P.W. Anderson, Phys. Rev. {\bf 109}, 1498 (1958).

\bibitem{7}  B.I. Shklovskii, B. Shapiro, B. Sears, P. Labrianides
and H.B. Shore, Phys. Rep. B {\bf 47}, 11487 (1993).

\bibitem{8}  I.K. Zharekeshev and B. Kramer, Phys. Rev. Lett.
{\bf 79}, 717 (1997).

\bibitem{9} E.B. Bogomolny, U. Gerland and C. Schmit, Phys. Rev. E
{\bf 59}, R1315 (1999) proposed the semi-Poisson
$P(S)$ for certain pesudointegrable billiards and pointed out its
similarity to the Anderson model at the transition point.

\bibitem{10}  G.N. Katomeris and S.N. Evangelou,
Eur. Phys. J. B{\bf 16}, 133 (2000).

\bibitem{11}  C. Gros,
Z. Phys. B -Condensed Matter {\bf 86}, 359 (1992).
 
\bibitem{12}  M. Kappus and F. Wegner,
 Z. Phys. B -Condensed Matter {\bf 45}, 15 (1981).
 
\bibitem{13}  H. Schomerus and M. Titov,
 Phys. Rev. B {\bf 67}, 100201(R) (2003) showed
a few percent deviations from single parameter scaling 
at special energies, e.g. at $E=0$.

\bibitem{14}  S.N. Evangelou and J.-L. Pichard,
 Phys. Rev. Lett. {\bf 84}, 1643 (2000).

\bibitem{15}  M. Batch, L. Schweitzer, I.Kh. Zharekeshev and B. Kramer,
 Phys. Rev. Lett. {\bf 77}, 1552 (1996).

\bibitem{16}  Y. Alhassid, 
 Rev. Mod. Phys. {\bf 72}, 845 (2000).
\end{thebibliography}
\end{document}